\documentclass{article}
\usepackage{amsfonts}

\usepackage{amsmath}

\newcommand{\bee}{\begin{equation}}
\newcommand{\ee}{\end{equation}}
\def\Ve{\mbox{\raise1.3pt\hbox{\tiny $\vert$}}}
\def\Vve{\mbox{\raise2pt\hbox{\scriptsize $\vert$}}}

\def\nie{\mbox{\kern1.5pt$\subset$\kern-5pt
\hbox{$\setminus$} }}

\input{tcilatex}

\begin{document}

\title{Quantum Stochastic Differential Equation\\
for Unstable Systems}
\author{V.\,P.~Belavkin \\
Mathematics Department, University of Nottingham\\
Nottingham NG7 2RD, United Kingdom \and P.~Staszewski \\
Institute of Mathematics, Pedagogical University of Bydgoszcz\\
85-072 Bydgoszcz, Poland}
\date{Received 10 December 1999\\
Published in: \textit{J. Math. Phys.} \textbf{41} No 11, 2000, 7220--7233.}
\maketitle

\begin{abstract}
A semi-classical non-Hamiltonian model of a spontaneous collapse of unstable
quantum system is given. The time evolution of the system becomes
non-Hamiltonian at random instants of transition of pure states to reduced
ones, $\eta \mapsto C\eta $, given by a contraction $C$. The counting
trajectories are assumed to satisfy the Poisson law. A unitary dilation of
the concractive stochastic dynamics is found. In particular, in the limit of
frequent detection corresponding to the large number limit we obtain the It%
\^{o}-Schr\"{o}dinger stochastic unitary evolution for the pure state of
unstable quantum system providing a new stochastic version of the quantum
Zeno effect.
\end{abstract}

\section{Introduction and summary}

The decay process is by its nature discontinuous and takes place at random
instants of time. Nevertheless, some authors succeeded in describing quantum
unstable systems by considering ``smoothed'' time evolution of unstable
systems in the dynamical semigroup approach.

The use of one parameter contracting semigroup in a Hilbert space \cite{1}
-- \cite{4} for the description of the dynamics of unstable quantum system $%
\mathcal{S}$ generalizes the law of exponential decay saying that the number
of particles in a given state which have not decayed up to $t$ is an
exponential function of time; $n(t)=n(0)\exp [-\lambda t]$\thinspace , $%
\lambda >0$\thinspace\ $t\geq 0$\thinspace . Let $\mathcal{H}$ be a Hilbert
space of $\mathcal{S}$, let $\psi (0)\in \mathcal{H}$ denotes an initial
(pure) state of $\mathcal{S}$\thinspace . It is assumed that for any $t\geq 0
$ the state of $\mathcal{S}$ is given by formula 
\begin{equation}
\psi (t)\;=\;V(t)\psi (0)\,,  \label{V}
\end{equation}%
where the family $\{V(t)\,,\,\,t\geq 0\}$ of bounded operators on $\mathcal{H%
}$ satisfies the following conditions: (a) $\parallel \!\!V(t)\!\!\parallel
\,\,\,\,\leq 1\,,\,\,t\geq 0$\thinspace , (b) $V(0)=I$\thinspace , (c) $%
V(t_{1}\!+\!t_{2})=V(t_{1})\,V(t_{2})\,,\,\,t_{1}\,,\,\,t_{2}\geq 0$%
\thinspace , (d) the map $t\mapsto V(t)$ is strongly continuous.

The state (\ref{V}) is normalized to the probability $p(t)\, =\,\,\,
\parallel \!\! \psi(t) \!\!\parallel$ of finding the system undecayed at $t$%
, moreover $p(t)$ monotonically decreases as the semigroup is contracting.

By virtue of Sz-Nagy theorem \cite{Sz} there is a unitary dilation of the
dynamics $V(t) $ on the Hilbert space $\mathcal{K} = \mathcal{H}\oplus 
\mathcal{K} $, where $\mathcal{K}$ denotes the Hilbert space of the products
of the decay.

Let us assume that the decay of the state of the unstable quantum system $%
\mathcal{S}$ is represented by completely positive map $\mathcal{I}:T(%
\mathcal{H})\rightarrow T(\mathcal{H})$ of the form \cite{Ozawa} 
\begin{equation}
\mathcal{I}\kern0.3pt\rho (t)\;=\;C\rho (t)C^{\ast }\,,\quad C^{\ast
}C\;\leq \;I\,,  \label{I}
\end{equation}%
where $I$ is the identity operator in $\mathcal{H}$, the Hilbert space of $%
\mathcal{S}$. Then the time evolution of the mixed state of the system in
question is given by strongly continuous contracting semigroup with the
generator of the form \cite{Dav1, Fonda} 
\begin{equation}
\displaystyle{\frac{\mathrm{d}\rho ^{\lambda }(t)}{\mathrm{d}t}}\;=\;-%
\mathrm{i}\,[H,\,\rho ^{\lambda }(t)]+\lambda \big(\mathcal{I}-I\big)\rho
^{\lambda }(t)\,,  \label{equ}
\end{equation}%
where $H$ denotes the hamiltonian of the unstable quantum system $\mathcal{S}
$, and $\lambda >0$ is the decay ratio. The mixed state $\rho ^{\lambda }(t)$
satisfying the dynamical evolution equation (\ref{equ}) is normalized to the
survival probability $\mbox{\rm
Tr\kern1.5pt}\rho ^{\lambda }(t)$ for which 
\begin{equation}
\displaystyle{\frac{\mathrm{d}}{\mathrm{d}t}}\kern.7pt\mbox{\rm Tr\kern1.5pt}%
\rho ^{\lambda }(t)\;=\;\mathrm{Tr}\big[(C^{\ast }C-I)\rho ^{\lambda }(t)%
\big]\;\leq \;0\,.  \label{1.4}
\end{equation}

In Sect.~2 we give a semi-classical non-Hamiltonian model of spontaneous
collapse of an unstable quantum system. The Hamiltonian time-evolution of
the system becomes non-Hamiltonian at random instants of transitions $\eta
\mapsto C\eta $ of pure states to reduced ones, given by the contraction $C$%
. It is assumed that the counting trajectories, consisted of instants of
occurrences of the collapse, are distributed according to the Poisson law.
We find the time-development of the classical state propagator $V_{t}$ in $%
\mathcal{H}$ in the form of It\^{o} stochastic equation with respect to the
classical Poisson process. Consequently, we obtain nonmixing It\^{o}
stochastic equations for pure (resp.~mixed) states of the unstable quantum
system $\mathcal{S}$. It is shown that the averaged density matrix
corresponding to the statistical mixture of collapsed states satisfies eq.~(%
\ref{equ}). Assuming that each collapse $\eta \mapsto C\eta $ slightly
changes the state of $\mathcal{S}$ ($I-C=\lambda ^{-1}R$ with bounded $R$
satisfying for large $\lambda $ the condition $R^{\ast }R\leq \lambda
(R+R^{\ast })$ ) we find the contracting semigroup equation resulting from
the stochastic dynamics in the large number limit $\lambda \rightarrow
\infty $.

In Sect.~3 we give the quantum stochastic representation $\widehat{V}_{t}$
of the classical stochastic propagator $V_{t}$ in $\mathcal{H}$ as an
operator-valued process in the Hilbert space $\mathcal{H}\otimes \mathcal{F}$%
, where $\mathcal{F}=F_{+}(L^{2}(\mathbb{R}_{+}))$ is the Bose Fock space
over the single-particle space of square-integrable complex functions on $%
\mathbb{R}_{+}$. To this end we employ the generating functional method
described in this section.

As a unitary dilation of a causal contractive cocycle $V_{t}$ in $\mathcal{H}
$ cannot in general be obtained from a causal unitary stochastic cocycle $%
U_{t}$ in the same Hilbert space $\mathcal{H}$, it is impossible to find a
Hamiltonian semiclassical dynamics giving the contractive stochastic
dynamics of the unstable quantum system as the reduced one. Therefore, we
consider the unitary dilation of the contraction $C$ in an extended Hilbert
space $\mathcal{H}\otimes \mathbb{C}^{2}$, the latter can be interpreted as
the Hilbert space of \textquotedblleft quantum meter\textquotedblright\
detecting the death or life of the unstable particle. The unitary dilation
of the contractive stochastic cocycle $V_{t}$, cf.~\cite{BelCMP97}, is then
realized as a causal unitary cocycle $U_{t}$ in a Hilbert tensor product $%
\mathcal{H}\otimes \mathcal{F}_{\bullet }$, where $\mathcal{F}_{\bullet }=%
\mathcal{F}_{+}(\mathbb{C}^{2}\otimes L^{2}(\mathbb{R}_{+}))$, the Bose Fock
space over one particle space $\mathbb{C}^{2}\otimes L^{2}(\mathbb{R}_{+})$,
Sect.~4. We consider two cases of the unitary dilation (4.1) $S$ of $C$ in $%
\mathcal{H}\otimes \mathbb{C}^{2}$: (a) with $S$ in the form of Hermitian
block-matrix (4.3-4),\break (b) non-Hermitian unitary block matrix (4.22).
In case (a) we find the QSDE for the unitary evolution in $\mathcal{H}%
\otimes \mathcal{F}_{\bullet }$ with respect to the quantum stochastic
Poisson matrix process of intensity $\lambda $. In case (b) we find the
limit (as $\lambda \rightarrow \infty $) of the unitary evolution using the
generating functional method described in Sect.~3. The limiting unitary
evolution in $\mathcal{H}\otimes \mathcal{F}_{\bullet }$ has the form of the
diffusion QSDE with respect to the field momentum process being quantum
stochastic representation of the standard Wiener process $w_{t}$ in the Fock
space of the representation of the Poisson process. Hence, we obtain (in the
representation in which the momentum process is diagonal) that the dilation
of the weakly random contractive process with the rate $\lambda \rightarrow
\infty $ is described by the It\^{o}-Schr\"{o}dinger equation for the pure
state (in $\mathcal{H}$) of the unstable system. The obtained result
provides a new \emph{stochastic} version of the \emph{quantum Zeno effect} 
\cite{Friedman, MisraSudarshan}, the limiting dynamics becomes reversible as
the reductions of decaying amplitude can be compensated by the field
fluctuations.

However, while in this paper we do not stress the problem of the Markovian
dynamics of a continuously observed (in time) quantum system (the state of
which undergoes the collapse), we would like to mention that this important
problem of quantum mechanics has been solved in the framework of quantum
stochastic calculus, cf.~\cite{Bel1, BelBa, BelCMP92, Sta, BelFound, AKS}
and the literature quoted therein.

\section{A stochastic model for an unstable quantum system}

\setcounter{equation}{0} Now we define a stochastic phenomenological model
of spontaneous collapse of an unstable quantum system. It is described as a
semiclassical non-Hamiltonian system with a Hilbert space $\mathcal{H}$ of
pure quantum states $\eta \in \mathcal{H}$, together with a classical
probability space of sequences $\omega =\{t_{1}\,,\,t_{2}\,,\,\ldots
\,\}\subset \mathbb{R}_{+}$ of the random time instants $t_{1}<t_{2}<\,%
\ldots \,$ of some events (reductions, transitions), which can demolish
eventually the quantum system. We shall assume that the sequences $\omega
\in \Omega $ are \textit{a priori\/} distributed according to the Poisson
law, given for each $t\in \mathbb{R}_{+}$ by the \textquotedblleft
input\textquotedblright\ probability measure $P_{t}^{\lambda }$ on the
measurable subsets of finite subsequences $\omega _{t}=\omega \cap \lbrack
\,0\,,\,t)$ as 
\begin{equation}
P_{t}^{\lambda }(\mathrm{d}\omega )\;=\;\lambda ^{n}\mathrm{e}^{-\lambda t}%
\mathrm{d}t_{1}\mathrm{d}t_{2}\cdot \,\ldots \,\cdot \mathrm{d}%
t_{n_{t}(\omega )}\,.
\end{equation}%
Here $n_{t}(\omega )=|\omega _{t}|$ is the random number of the events up to
time $t$, $\lambda \geq 0$ is the intensity of the stationary Poisson
process $t\mapsto n_{t}$, i.e. the average number of the events per unit of
time. 
The probability of $n$ events, on each interval $[r\,,\,r+t)$, is given by
the Poissonian distribution 
\begin{equation}
p_{t}^{\lambda }(n)\;=\;\int\limits_{r\leq t_{1}<\,\ldots
}\int\limits_{<t_{n}<t+r}P_{t}^{\lambda }(\mathrm{d}\omega )\;=\;%
\displaystyle{\frac{(\lambda t)^{n}}{n!}}\mathrm{e}^{-\lambda t}\,,
\label{Poisson}
\end{equation}%
independently of $r\in \mathbb{R}_{+}$.

Each event $t \in \omega$ results in an instantaneous change (collapse) $%
\eta \mapsto C \eta$ of the state of the quantum system, mapping a
normalized state $\eta \in \mathcal{H}\,, \,\, \parallel \! \eta \!
\parallel^{2} = \langle \eta \, | \, \eta \rangle = 1$ to the reduced state $%
C \eta$ with the survival probability $\parallel \!\! C \kern.2pt \eta \!\!
\parallel^{2} \leq 1$. This change satisfies quantum superposition principle
i.e. it is described by a linear contraction $C: \mathcal{H} \rightarrow 
\mathcal{H}$, $C^{\ast}C \leq I$. The case $C^{\ast}C =I$ of isometric $C$
corresponds to a stable (in the positive direction of time) quantum
stochastic evolution, with the survival probability one for each state $\eta
\in \mathcal{H}$.

If we assume that the quantum system between the reductions is conservative
and Hamiltonian, then the nonmixing stochastic evolution $\eta \mapsto \chi
_{t}(\omega )$ of the initial quantum normalized states $\eta $ to the pure
states $\chi _{t}(\omega )\in \mathcal{H}$ is defined by the measurable maps 
$\chi _{t}:\Omega \rightarrow \mathcal{H}$ as 
\begin{equation}
\chi _{t}\;=\;V_{t}\eta \,,  \label{pure}
\end{equation}%
where 
\begin{equation}
V_{t}(\omega )\;=\;\mathrm{e}^{\mathrm{i}H(t_{n_{t}(\omega )}-t)}C\,\ldots \,%
\mathrm{e}^{\mathrm{i}H(t_{1}-t_{2})}C\mathrm{e}^{-\mathrm{i}%
Ht_{1}}\;=\;V_{t}(t_{1},\,\ldots \,,t_{n})\,.  \label{product}
\end{equation}%
Here $\{\mathrm{e}^{-\mathrm{i}Ht}\,,\,t\in \mathbb{R}_{+}\}$ is a strongly
continuous group of unitary operators with a selfadjoint generator $H$ (the
hamiltonian of the quantum system in the units $\hbar =1$), and for each $%
t<\infty $ the product (\ref{product}) is finite as $n_{t}(\omega )<\infty $
with probability one. Hence, the stochastic propagator $V_{t}(\omega )$ is
well-defined as a contraction in $\mathcal{H}$, giving for each $\omega \in
\Omega $ the monotonically decreasing probability of the unstable particle
at the time $t$, 
\begin{equation}
\parallel \!\chi _{t}(\omega )\!\parallel ^{2}\;=\;\langle V_{t}(\omega
)\eta \,|\,V_{t}(\omega )\eta \rangle \;\leq \,\,\;\parallel \!\chi
_{r}(\omega )\!\parallel ^{2}\;\leq \;1\,,\quad \forall r\in \lbrack
0\,,\,t)\,.
\end{equation}%
Thus, the survival probability $q_{t}(\omega )=\,\,\parallel \!\!\chi
_{t}(\omega )\!\!\parallel ^{2}$ is obtained as a positive decreasing
stochastic process with the initial value $q_{0}(\omega )=1$. Its
expectation gives a deterministic monotonically decreasing process of the
averaged survival probability 
\begin{equation}
q^{\lambda }(t)\;=\;\int q_{t}(\omega )\,P_{t}^{\lambda }(\mathrm{d}\omega
)\;\leq \;q^{\lambda }(r)\;\leq \;1\,,\quad \forall r\in \lbrack 0,t)\,.
\end{equation}%
The stochastic process $q_{t}(\omega )$ defines quantum \textit{a
posteriori\/} states \cite{Ozawa, BelBa} of the nondemolished quantum system
by 
\begin{equation}
\eta _{t}(\omega )\;=\;\chi _{t}(\omega )/\parallel \!\chi _{t}(\omega
)\!\parallel \,,\quad \forall \omega :\,\,q_{t}(\omega )\;\neq \;0\,,
\end{equation}%
and the output statistics of the finite sequences $\omega _{t}\subset
\lbrack 0\,,\,t)$. The latter is given together with the probability of the
survival event of the quantum system at the time $t$ by the output
probability measure $Q_{t}^{\lambda }($\textrm{d}$\omega )=q_{t}(\omega
)P_{t}^{\lambda }($\textrm{d}$\omega )$, normalized to the probability $%
q^{\lambda }(t)$. The averaged density matrix 
\begin{equation}
\rho ^{\lambda }(t)\;=\;\int \rho _{t}(\omega )P_{t}^{\lambda }(\mathrm{d}%
\omega )\;=\;\int \eta _{t}(\omega )\eta _{t}(\omega )^{\ast }Q_{t}^{\lambda
}(\mathrm{d}\omega )\,,
\end{equation}%
corresponding to the statistical mixture of the collapsed states 
\begin{equation}
\rho _{t}(\omega )\;=\;\chi _{t}(\omega )\chi _{t}(\omega )^{\ast }
\label{stodens}
\end{equation}%
by the time $t$ satisfies equation (\ref{equ}) 
\begin{equation}
\displaystyle{\frac{\mathrm{d}\rho ^{\lambda }(t)}{\mathrm{d}t}}\;=\;-%
\mathrm{i}\,[H\,,\,\rho ^{\lambda }(t)]\,+\,\lambda \big(C\rho ^{\lambda
}(t)C^{\ast }-\rho ^{\lambda }(t)\big)\,,\quad \rho ^{\lambda }(0)\;=\;\eta
\eta ^{\ast }\,.
\end{equation}%
Indeed, this equation can be resolved by the Dyson-von Neumann series \cite%
{Dav1} 
\begin{equation}
\rho ^{\lambda }(t)\;=\;\sum\limits_{n=0}^{\infty }\,\lambda
^{n}\!\!\int\limits_{0\leq t_{1}<\ldots
}\int\limits_{<t_{n}<t}\!\!V_{t}(t_{1},\,\ldots \,,t_{n})\kern.5pt\sigma
V_{t}(t_{1},\,\ldots \,,t_{n})^{\ast }\mathrm{e}^{-\lambda t}\,\mathrm{d}%
t_{1}\,\ldots \,\mathrm{d}t_{n},
\end{equation}%
which for $\sigma =\eta \eta ^{\ast }$ is the mean value of the stochastic
density matrix $\rho _{t}(\omega )$ with respect to the Poisson probability
measure (\ref{Poisson}). Thus, the averaged dynamics $\sigma \mapsto \rho
^{\lambda }(t)$ for the unstable system is continuous, contractive 
\begin{equation*}
\displaystyle{\frac{\mathrm{d}}{\mathrm{d}t}}\kern.7pt\mbox{\rm Tr\kern1.5pt}%
\rho ^{\lambda }(t)\;=\;\mathrm{Tr}\big[(C^{\ast }C-I)\rho ^{\lambda }(t)%
\big]\;\leq \;0\,,
\end{equation*}%
being normalized to the survival probability $q^{\lambda }(t)=\mathrm{Tr}%
\rho ^{\lambda }(t)$, and mixing.

However, the nonmixing stochastic dynamics 
\begin{equation}
\sigma \mapsto \rho _{t}(\omega )\;=\;V_{t}(\omega )\sigma V_{t}(\omega
)^{\ast }\,,
\end{equation}%
which can be studied in terms of Hilbert space propagators $V_{t}(\omega ):%
\mathcal{H}\rightarrow \mathcal{H}$ is discontinuous and cannot be defined
by a differential evolution equation in an ordinary sense. Indeed, the
stochastic propagator $V_{t}(\omega )$ is strongly right discontinuous at
the points of the collapse $t\in \omega $, but it has strong limits at each $%
t\in \mathbb{R}$. It is strongly continuous from the left, satisfying the
usual Schr\"{o}dinger equation in terms of the left differentials \textrm{d}$%
_{-}V_{t}=V_{t}-V_{t-\mathrm{d}t}=-$\textrm{i}$HV_{t}$. However, the Schr%
\"{o}dinger equation does not recover the stochastic propagator $V_{t}$ but
only its nonstochastic unitary part $\mathrm{e}^{-\mathrm{i}Ht}$.

The proper differential equation for $V_{t}$ can be written as the
stochastic equation in It\^{o} sense 
\begin{equation}
\mathrm{d}V_{t}(\omega )+\mathrm{i}HV_{t}(\omega )\,\mathrm{d}%
t\;=\;(C-I)V_{t}(\omega )\,\mathrm{d}n_{t}(\omega )\,,\quad V_{0}(\omega
)\;=\;I\,.  \label{stochastic}
\end{equation}%
Here \textrm{d}$V_{t}$ is forward or symmetric or any other increment of $%
V_{t}$ but not the backward differential \textrm{d}$_{-}V_{t}$ for which 
\textrm{d}$_{-}n_{t}(\omega )=n_{t}(\omega )-n_{t-\mathrm{d}t}(\omega )=0$
for all $\omega \in \Omega $ \big(\textrm{d}$n_{t}(\omega )=|\omega \cap
\lbrack t-$\textrm{d}$t\,,\,t)|$ is zero as soon as \textrm{d}$%
t<t_{n+1}-t_{n}$ for $n=n_{t}(\omega )$\big).

To be definite, we shall always assume that \textrm{d}$V_{t}$ (and
respectively \textrm{d}$n_{t}$) is the forward differential $V_{t+\mathrm{d}%
t}-V_{t}$ and \textrm{d}$n_{t}(\omega )=|\omega \cap \lbrack t\,,\,t+$%
\textrm{d}$t)|$ is either zero (if $t\notin \omega $) or one (if $t\in
\omega )$) for a sufficiently small \textrm{d}$t$ $\big($\textrm{d}$%
t<t_{n+1}-t_{n}$ for $n=n_{t}(\omega )\big)$. Thus the stochastic equation (%
\ref{stochastic}) coincides with the Schr\"{o}dinger equation at when there
is no collapse, $t\not\in \omega $, and \textrm{d}$V_{t}(\omega
)=(C-I)V_{t}(\omega )$ at the points of collapse corresponding to the
reduction $V_{t+0}(\omega )=CV_{t}(\omega )$ at $t\in \omega $ and \textrm{d}%
$t\rightarrow 0$. One can prove that the stochastic equation (\ref%
{stochastic}) has only one solution, (\ref{product}). From (\ref{stochastic}%
) and (\ref{pure}) one obtains the stochastic equation for the pure state $%
\chi _{t}$ 
\begin{equation}
\mathrm{d}\chi _{t}(\omega )\;=\;-\mathrm{i}H\chi _{t}(\omega )\,\mathrm{d}%
t+(C-I)\chi _{t}(\omega )\,\mathrm{d}n_{t}(\omega )\,,\quad \chi _{0}(\omega
)\;=\;\eta \,.  \label{chi}
\end{equation}%
The stochastic density matrix (\ref{stodens}) can also be obtained by
iterations as the unique solution to the stochastic differential equation 
\begin{equation}
\mathrm{d}\rho _{t}\;=\;-\mathrm{i}[H\,,\,\rho _{t}]\,\mathrm{d}t+(C\rho
_{t}C^{\ast }-\rho _{t})\,\mathrm{d}n_{t}\,,\quad \rho _{0}(\omega
)\;=\;\sigma \,.  \label{storho}
\end{equation}%
Note, that this equation coinciding with the von Neumann equation, \textrm{d}%
$\rho _{t}/$\textrm{d}$t=-$\textrm{i}$[H,\,\rho _{t}]$ at $t\not\in \omega $
and with \textrm{d}$\rho _{t}=C\rho _{t}C^{\ast }-\rho _{t}$ at the points
of the collapse $\rho _{t+0}(\omega )=C\rho _{t}(\omega )C^{\ast
}\,,\,\,t\in \omega $, can be derived from the stochastic equation (\ref{chi}%
). Indeed, by virtue of the It\^{o} differentiation formula applied to the
product $\chi _{t}\chi _{t}^{\ast }$: 
\begin{equation}
\mathrm{d}(\chi _{t}\chi _{t}^{\ast })\;=\;\mathrm{d}\chi _{t}\cdot \chi
_{t}^{\ast }+\chi _{t}\cdot \mathrm{d}\chi _{t}^{\ast }+\mathrm{d}\chi
_{t}\cdot \mathrm{d}\chi _{t}^{\ast }  \label{Itof}
\end{equation}%
and the It\^{o} multiplication table 
\begin{equation}
(\mathrm{d}t)^{2}\;=\;0\,,\quad (\mathrm{d}n_{t})^{2}\;=\;\mathrm{d}%
n_{t}\,,\quad \mathrm{d}n_{t}\,\mathrm{d}t\;=\;0\;=\;\mathrm{d}t\,\mathrm{d}%
n_{t}
\end{equation}%
one easily obtains (2.15). Then, the averaged mixing equation for $\rho
_{t}^{\lambda }$ is obtained from (\ref{storho}) by formal replacement 
\textrm{d}$n_{t}$ with $\lambda $\textrm{d}$t$ corresponding to the averaged
number $n_{t}^{\lambda }=\lambda t$ for the Poisson process with the
intensity $\lambda $.

The strongly continuous nonmixing evolution 
\begin{equation}
\rho (t)\;=\;\mathrm{e}^{-Kt}\sigma \mathrm{e}^{-K^{\ast }t}\,,  \label{rhot}
\end{equation}%
with 
\begin{equation}
K\;=\;\mathrm{i}H+\lambda (I-C)  \label{K}
\end{equation}%
corresponding to equation (\ref{equ}) follows from (2.15) in the \emph{large
number limit\/} $\lambda \rightarrow \infty $ of the stochastic evolution
under the condition that each collapse $\eta \mapsto C\eta $ only slightly
changes the state of the unstable system such that $\eta -C\eta $ is
inversely proportional to $\lambda $. Indeed, substituting in equations (\ref%
{stochastic}), (\ref{storho}) $I-C$ by $\lambda ^{-1}R$, where $R$ satisfies
the condition $R^{\ast }R\leq \lambda (R+R^{\ast })$ for large $\lambda $,
we obtain 
\begin{equation}
\mathrm{d}V_{t}(\omega )+(R\lambda ^{-1}\,\mathrm{d}n_{t}(\omega )+\mathrm{i}%
H\,\mathrm{d}t)V_{t}(\omega )\;=\;0\,.
\end{equation}%
As in the large number limit $\lambda ^{-1}n_{t}(\omega )$ converges to $t$
with probability one, this dynamics becomes nonstochastic, satisfying the
ordinary differential equation 
\begin{equation}
\frac{\mathrm{d}}{\mathrm{d}t}V(t)+KV(t)\;=\;0\,,\qquad V(0)=I\,,
\label{diff}
\end{equation}%
for $V(t)=\lim_{\lambda \rightarrow \infty }V_{t}=V_{t}^{0}$ with $K=$%
\textrm{i}$H+R$. It has a unique strongly continuous solution $\mathrm{e}%
^{-Kt}$, which is a semigroup of contractions as $K+K^{\ast }=R+R^{\ast
}\geq 0$. The corresponding nonmixing equation 
\begin{equation}
\frac{\mathrm{d}}{\mathrm{d}t}\rho (t)+K\rho (t)+\rho (t)K^{\ast }\;=\;0
\label{nonmixing}
\end{equation}%
for nonstochastic density matrix $\rho (t)=V(t)\sigma V^{\ast }(t)$ can be
obtained in the limit $\lambda \rightarrow \infty $ from (2.10), or directly
from It\^{o} equation (\ref{storho}) with $C=I-\lambda ^{-1}R$. This is not
surprising as the large number limit coincides with its average, thus
becoming nonmixing in this limit.

\section{A generating functional method and quantum stochastic representation%
}

\setcounter{equation}{0} A very convenient method of treating stochastic
equations is based on studying the corresponding generating functional
equations. The generating functional for a causal stochastic process $\chi
_{t}(\omega )$ obtained by solving a stochastic equation with respect to the
Poisson process of the intensity $\lambda $ is defined as the averaged
product $\chi _{t}^{f}=\chi _{t}\varepsilon _{t}^{f}$, 
\begin{equation}
\breve{\chi}_{t}(f)\;=\;\langle \chi _{t}^{f}\rangle \;:=\;\int \chi
_{t}(\omega )\varepsilon _{t}^{f}(\omega )\,P_{t}^{\lambda }(\mathrm{d}%
\omega )\,,  \label{3.1}
\end{equation}%
where $\varepsilon _{t}^{f}(\omega )$ is the stochastic exponent for the
martingale process $m_{t}=n_{t}-\lambda t$, satisfying the stochastic
equation 
\begin{equation}
\lambda ^{1/2}\mathrm{d}\varepsilon _{t}^{f}(\omega )\;=\;f(t)\varepsilon
_{t}^{f}(\omega )\,\mathrm{d}m_{t}(\omega )\,,\quad \varepsilon
_{0}^{f}(\omega )\;=\;1\,.  \label{epsilon}
\end{equation}%
Here $f(t)$ is a nonstochastic complex locally integrable test function such
that $|1+\lambda ^{-1/2}f(t)|\leq 1$ for all $t$. The solution to this
stochastic equation can be written as 
\begin{equation}
\varepsilon _{t}^{f}(\omega )\;=\;\exp [-\lambda ^{1/2}\int_{0}^{t}f(r)\,%
\mathrm{d}r]\prod\limits_{r\in \omega _{t}}(1+\lambda ^{-1/2}f(r))\,,
\label{eprep}
\end{equation}%
where $\omega _{t}=\omega \cap \lbrack 0\,,\,t)$. The inverse transform $%
\breve{\chi}_{t}\mapsto \chi _{t}$ can be written in terms of the series of
iterated stochastic integrals 
\begin{equation}
\int \lambda ^{-|\tau |/2}\varphi (\tau )\,\mathrm{d}m_{\tau
}:=\sum\limits_{n=0}^{\infty }\,\,\lambda ^{-n/2}\int\limits_{0\leq
r_{1}<\ldots }\int\limits_{<r_{n}<\infty }\varphi (r_{1}\,,\,\ldots
\,,\,r_{n})\,\mathrm{d}m_{r_{1}}\,\ldots \,\mathrm{d}m_{r_{n}}\,,
\label{iter}
\end{equation}%
as $\chi _{t}(\omega )=\int \lambda ^{-|\tau |/2}\widetilde{\chi }_{t}(\tau
)\,$\textrm{d}$m_{\tau }$, where $\widetilde{\chi }_{t}(r_{1}\,,\,\ldots
\,,\,r_{n})$ are the functional derivatives of $\breve{\chi}_{t}(f)$ with
respect to $f(r_{1})$, $f(r_{2})\,,\,\ldots \,,\,f(r_{n})$; 
\begin{equation}
\widetilde{\chi }_{t}(r_{1}\,,\,\ldots \,,\,r_{n})\;=\;\delta ^{n}\breve{\chi%
}_{t}(f)/\delta f(r_{1})\,\ldots \,\delta f(r_{n})|_{f=0}\,.
\label{functder}
\end{equation}

In particular, the stochastic exponent $\varepsilon _{t}^{g}$ has the
exponential generating functional 
\begin{equation}
\breve{\varepsilon}_{t}^{g}(f)\;=\;\int \varepsilon _{t}^{f}(\omega
)\varepsilon _{t}^{g}(\omega )\,P_{t}^{\lambda }(\mathrm{d}\omega )\;=\;%
\mathrm{e}^{\int_{0}^{t}g(r)f(r)\,\mathrm{d}r}  \label{expgenfun}
\end{equation}%
such that $\breve{\varepsilon}_{t}^{g}(f)=\breve{\varepsilon}_{t}^{f}(g)$.
Indeed, it follows from the multiplication formula for stochastic exponents, 
\begin{equation}
\varepsilon _{t}^{f}\varepsilon _{t}^{g}\;=\;\varepsilon _{t}^{f\dot{+}g}%
\mathrm{e}^{\int_{0}^{t}g(r)f(r)\,\mathrm{d}r}\,,  \label{3.7}
\end{equation}%
where 
\begin{equation}
f\!\dot{+}\!g=f+\lambda ^{-1/2}fg+g
\end{equation}%
and $\langle \varepsilon _{t}^{f\dot{+}g}\rangle =1$ as it is easily seen in
the explicit representation (\ref{eprep}). Note that $\varepsilon _{t}^{g}$
can be written in the form of the multiple integral (\ref{iter}) as $%
\varepsilon _{t}^{g}=\varepsilon ^{g_{t}}$, 
\begin{equation}
\varepsilon ^{g}\;=\;\sum\limits_{n=0}^{\infty }\,\lambda
^{-n/2}\,\int\limits_{0\leq r_{1}<\ldots }\int\limits_{<r_{n}<\infty
}g(r_{1})\,\ldots \,g(r_{n})\,\mathrm{d}m_{r_{1}}\,\ldots \,\mathrm{d}%
m_{r_{n}}\,,  \label{epsilong}
\end{equation}%
where $g_{t}(r)=g(r)$, $r<t$ and $g_{t}(r)=0$, $r\geq t$. This follows from $%
\breve{\varepsilon}_{t}^{g}(f)=\breve{\varepsilon}^{g_{t}}(f)$, where $%
\breve{\varepsilon}^{g}(f)=\exp \{\int_{0}^{\infty }g(r)f(r)\,$\textrm{d}$r\}
$ corresponds to the kernel 
\begin{equation}
\widetilde{\varepsilon }^{g}(r_{1},\,\ldots \,,\,r_{n})\;=\;g(r_{1})\,\ldots
\,g(r_{n})\,.  \label{tildetrans}
\end{equation}%
Note, that the Hilbert space $L_{P}^{2}(\Omega )$ of complex random
functions $\chi (\omega )$ with $\int_{\Omega }|\chi (\omega )|^{2}\,P($%
\textrm{d}$\omega )<\infty $ is isomorphic to the Fock space of their
transforms $\widetilde{\chi }$ with respect to the scalar product 
\begin{equation}
(\varphi |\widetilde{\chi })=\sum\limits_{n=0}^{\infty
}\,\,\,\int\limits_{0\leq r_{1}<\ldots }\int\limits_{<r_{n}<\infty }\!\!\!%
\bar{\varphi}(r_{1}\,,\,\ldots \,,\,r_{n})\widetilde{\chi }(r_{1}\,,\,\ldots
\,,\,r_{n})\,\mathrm{d}r_{1}\,\ldots \,\mathrm{d}r_{n}\;.  \label{scalarpr}
\end{equation}%
Thus, the generating functional (\ref{3.1}) can be written in terms of the
scalar product (\ref{scalarpr}) as follows 
\begin{equation}
\breve{\chi}_{t}(\bar{g})\;=\;(\widetilde{\varepsilon }_{t}^{g}|\widetilde{%
\chi }_{t})\;=\;\int\limits_{\tau \subset \lbrack t,0)}\overline{\widetilde{%
\varepsilon }_{t}^{g}}(\tau )\widetilde{\chi }_{t}(\tau )\,\mathrm{d}\tau 
\label{genfu31}
\end{equation}%
for the tilde transform (\ref{tildetrans}) of (\ref{epsilong}) and $%
\widetilde{\chi }_{t}$.

Let us now obtain a differential equation for the generating functional $%
\breve{\chi}_{t}$ of the stochastic process $\chi _{t}$, satisfying the
equation (\ref{chi}). By differentiating the pointwise product $\chi
_{t}^{f}(\omega )=\chi _{t}(\omega )\varepsilon _{t}^{f}(\omega )$ we obtain
the stochastic equation 
\begin{equation}
\mathrm{d}\chi _{t}^{f}+\big(\lambda ^{1/2}f(t)+\mathrm{i}H\big)\chi
_{t}^{f}\,\mathrm{d}t\;=\;\big(C(1+\lambda ^{-1/2}f(t))-I\big)\chi _{t}^{f}\,%
\mathrm{d}n_{t}\,,  \label{chief}
\end{equation}%
from (\ref{chi}) and (\ref{epsilon}) by applying the It\^{o} formula 
\begin{eqnarray}
\!\!\!\!\mathrm{d}\big(\chi _{t}\varepsilon _{t}^{f}\big)\!\! &=&\!\!\mathrm{%
d}\chi _{t}\cdot \varepsilon _{t}^{f}+\chi _{t}\cdot \mathrm{d}\varepsilon
_{t}^{f}+\mathrm{d}\chi _{t}\cdot \mathrm{d}\varepsilon _{t}^{f}
\label{Itofor} \\
\!\! &=&\!\![(C\!-\!I)\mathrm{d}n_{t}\!-\!\mathrm{i}H\mathrm{d}t\!+\!\lambda
^{-1/2}f(t)\mathrm{d}m_{t}\!+\!(C\!-\!I)\lambda ^{-1/2}f(t)\mathrm{d}%
n_{t}]\chi _{t}^{f}\,.  \notag
\end{eqnarray}

Thus, the generating functional $\breve{\chi}_{t}(f)=\langle \chi
_{t}^{f}\rangle $ satisfies the ordinary differential equation 
\begin{equation}
\displaystyle{\frac{\mathrm{d}}{\mathrm{d}t}}\breve{\chi}_{t}+\mathrm{i}H%
\breve{\chi}_{t}\;=\;(C-I)(\lambda ^{-1/2}f(t)+1)\lambda \breve{\chi}_{t}
\label{gechi}
\end{equation}%
with the initial condition $\breve{\chi}_{0}(f)=\eta $ for all $f$. The
increment \textrm{d}$n_{t}$ is replaced in (\ref{gechi}) by its average $%
\langle $\textrm{d}$n_{t}\rangle =\lambda $\textrm{d}$t$ because it does not
depend on $\chi _{t}$. The solution to this equation can be written in terms
of time ordered exponents $\breve{\chi}_{t}=\overset{\longleftarrow }{\exp }%
[-\int_{0}^{t}K^{\lambda }(r)\,$\textrm{d}$r]\eta $ as follows 
\begin{equation}
\breve{\chi}_{t}\;=\;\sum\limits_{n=0}^{\infty }(-1)^{n}\int\limits_{0\leq
r_{1}<\ldots }\int\limits_{<r_{n}<t}K^{\lambda }(r_{n})\,\ldots \,K^{\lambda
}(r_{1})\eta \,\mathrm{d}r_{1}\,\ldots \,\mathrm{d}r_{n}\,,  \label{solgechi}
\end{equation}%
where 
\begin{equation}
K^{\lambda }(r)\;=\;\lambda (I-C)(I+\lambda ^{-1/2}f(r))+\mathrm{i}H\,.
\label{Konlambda}
\end{equation}%
Thus, the tilde transform $\widetilde{\chi }_{t}$ of the stochastic function 
$\chi _{t}$ is given by 
\begin{equation}
\widetilde{\chi }_{t}(r_{1}\,,\,\ldots \,,\,r_{n})\;=\;\mathrm{e}%
^{(r_{n}-r)K}(C-I)\,\ldots \,\mathrm{e}^{(r_{1}-r_{2})K}(C-I)\mathrm{e}%
^{-r_{1}K}\eta \,,
\end{equation}%
where $K$ is given by formula (\ref{K}).

It is particularly simple to obtain the large number limit in terms of the
generating functional, one has $\breve{\chi}_{t}(f)\rightarrow \mathrm{e}%
^{-Kt}\eta $ as $\lambda \rightarrow \infty $ under the condition $\lambda
(I-C)\rightarrow R$, since obviously $K^{\lambda }(t)\rightarrow R+$\textrm{i%
}$H$.

It is well known \cite{HudsonParthasarathy} that the classical stochastic
Poisson process $n_{t}(\omega )$ has a quantum field representation $N_{t}=%
\widehat{n}_{t}$ in the Bosonic Fock space $\mathcal{F}$ over the single
quantum space $L^{2}(\mathbb{R}_{+})$ of square-integrable complex functions
on $\mathbb{R}_{+}$ in terms of the basic quantum stochastic processes of
number $\Lambda _{t}$, creation $A_{t}^{\ast }$, and annihilation $A_{t}$ on
the interval $[0\,,\,t)$. Let us also find the corresponding quantum
stochastic representation for the stochastic process $\chi _{t}$ satisfying
the equation (\ref{chi}).

Realizing $\mathcal{F}$ as the space of square-integrable summable functions 
$\varphi $ of the finite, ordered sequences $\tau =(r_{1}\,,\,\ldots
\,,\,r_{n})\,,\,\,\,r_{1}<\,\ldots \,<r_{n}\,$, 
\begin{equation}
\parallel \!\varphi \!\parallel ^{2}\;=\;\sum\limits_{n=0}^{\infty
}\,\,\,\int\limits_{0\leq r_{1}<\ldots }\int\limits_{<r_{n}<\infty }|\varphi
(r_{1}\,,\,\ldots \,,\,r_{n})|^{2}\mathrm{d}r_{1}\,\ldots \,\mathrm{d}%
r_{n}<\infty \,,
\end{equation}%
we can represent the canonical operator processes $A_{t}$, $A_{t}^{\ast }$, $%
\Lambda _{t}$ as 
\begin{equation}
A_{t}\varphi (\tau )=\int_{0}^{t}\dot{\varphi}(\tau ,r)\,\mathrm{d}r\,,\quad
A_{t}^{\ast }\varphi (\tau )=\sum_{r\in \tau }\varphi (\tau \setminus
r)\,,\quad \Lambda _{t}\varphi (\tau )=|\tau |\,\varphi (\tau )\,.
\end{equation}%
Here $n_{t}=|\tau _{t}|$ is the length of a subsequence $t_{1}\,,\,\ldots
\,,\,t_{n_{t}}<t$ of the sequence $\tau $ with $t_{n_{t+1}}\geq t$, $\tau
\setminus r$ is the subsequence without an element $r\in \tau $, and $\dot{%
\varphi}(\tau ,r)=\varphi (\tau \sqcup r)$, where $\tau \sqcup r$ is the
ordered sequence with an additional element $r\not\in \tau $.

Now, one can define the operator-valued representation $M_{t}=\widehat{m}_{t}
$ of the stochastic processes $m_{t}=n_{t}-\lambda t$ by the sum 
\begin{equation}
M_{t}\;=\;\Lambda _{t}+\sqrt{\lambda }\big(A_{t}+A_{t}^{\ast }\big)\;.
\label{procesM}
\end{equation}%
Any regular quantum stochastic process $X_{t}$ which is adapted with respect
to the family of commuting selfadjoint operators $\{M_{t}\,,\,\,t\in \mathbb{%
R}_{+}\}$ in $\mathcal{F}$ is given by the series of iterated integrals 
\begin{equation}
X_{t}\;:=\;\sum\limits_{n=0}^{\infty }\,\,\,\lambda
^{-n/2}\int\limits_{0\leq r_{1}<\ldots }\int\limits_{<r_{n}<t}\widetilde{%
\chi }(r_{1}\,,\,\ldots \,,\,r_{n})\,\mathrm{d}M_{r_{1}}\,\ldots \,\mathrm{d}%
M_{r_{n}}\,.  \label{Mintegral}
\end{equation}%
The map $\widetilde{\chi }_{t}\mapsto X_{t}$ is one-to-one because the
kernel $\widetilde{\chi }_{t}$ in (\ref{Mintegral}) is uniquely defined as
the image $\breve{X}_{t}:=X_{t}\varphi _{0}$ of $X_{t}=\int \widetilde{\chi }%
(\tau )\,$\textrm{d}$M_{\tau }$ on the vacuum state $\varphi _{0}(\tau
)=\delta _{0}^{|\tau |}$ ($\varphi _{0}$ is equal to zero if $n=|\tau |\neq 0
$ and is equal to one if $\tau =\emptyset $). If the kernel $\widehat{\chi }%
_{t}$ is given by the functional derivatives (\ref{functder}) of the
functional $\breve{\chi}_{t}$, (\ref{Mintegral}) can be formally written as
the normally ordered causal expression $X_{t}\,=\,\,:\!\breve{\chi}%
_{t}(\lambda ^{-1}\dot{M})\!\!:$ of the quantum field $\widehat{f}=\lambda
^{-1}\dot{M}_{t}$, where $\dot{M}$ is the generalized time derivative of (%
\ref{procesM}). The composition of the map $\breve{\chi}_{t}\mapsto X_{t}$
with the map $\chi _{t}\mapsto \breve{\chi}_{t}$ in (\ref{3.1}) defines an
operator representation $\chi _{t}\mapsto X_{t}$ called the quantum
stochastic representation of the process $\chi _{t}$. In particular, the
Wick exponent 
\begin{equation}
W_{t}^{g}\;=\;\int\limits_{\tau \subset \lbrack 0,t)}\prod\limits_{r\in \tau
}\big(g(r)/\lambda ^{1/2}\big)\mathrm{d}M_{\tau }\;=\;\widehat{\varepsilon }%
_{t}^{g}
\end{equation}%
defined as the unique solution to the operator differential equation 
\begin{equation}
\lambda ^{1/2}\mathrm{d}W_{t}^{f}\;=\;f(t)W_{t}^{f}\,\mathrm{d}M_{t}\,,\quad
W_{0}^{f}\;=\;\widehat{1}\,,
\end{equation}%
in terms of forward differentials \textrm{d}$M_{t}=M_{t+\mathrm{d}t}-M_{t}$,
is the quantum stochastic integral representation (\ref{eprep}) of the
solution to the stochastic differential equation (\ref{epsilon}) with the
tilde transform $\widetilde{\varepsilon }_{t}^{f}(\tau )=W_{t}^{f}\varphi
_{0}$. It has the operator multiplication 
\begin{equation}
W_{t}^{f}W_{t}^{g}\;=\;W_{t}^{f\dot{+}g}\,\exp \{\int_{0}^{t}f(r)g(r)\,%
\mathrm{d}r\}  \label{opmultiplication}
\end{equation}%
representing the stochastic multiplication (\ref{expgenfun}), and can be
formally written as the normally ordered exponent of $\lambda
^{-1/2}\int\limits_{0}^{t}g(r)\,$\textrm{d}$M(r)$ having the Wick symbol (%
\ref{expgenfun}). From this it follows that 
\begin{equation}
\big(W_{t}^{g}\varphi _{0}|X_{t}\varphi _{0}\big)\;=\;\big(W_{t}^{g}\varphi
_{0}|X_{t}\varphi _{0}\big)\;=\;\langle \widetilde{\varepsilon }_{t}^{g}|%
\widetilde{\chi }_{t}\rangle \;=\;\breve{\chi}_{t}(\bar{g})\,,
\end{equation}%
where $\breve{\chi}_{t}$ is the generating functional of a causal stochastic
process with the tilde transform $\tilde{\chi}_{t}$. Thus the generating
functional $\big(X_{t}^{f}\big)=\big(\varphi _{0}|X_{t}^{f}\varphi _{0}\big)$%
, defined for the operator integral $X_{t}=\widehat{\chi }_{t}$ as the
vacuum expectation of the commuting products $X_{t}^{f}=X_{t}W_{t}^{f}$,
coincides with the generating functional for the classical stochastic
process (\ref{iter}). This also proves the statistical equivalence of the
classical process $\chi _{t}$ and the quantum process $X_{t}=\widehat{\chi }%
_{t}$, having the kernel $\widetilde{\chi }_{t}=\breve{X}_{t}$ as the tilde
transform of $\chi _{t}$.

Now, we can define a quantum stochastic representation $\widehat{V}_{t}$ of
classical stochastic propagator $V_{t}(\omega )$ in $\mathcal{H}$ as an
operator-valued process acting in the Hilbert product $\mathcal{H}\otimes 
\mathcal{F}$ by the quantum stochastic differential equation 
\begin{equation}
\mathrm{d}\widehat{V}_{t}+\mathrm{i}H\widehat{V}_{t}\,\mathrm{d}t\;=\;(C-I)%
\widehat{V}_{t}\,\mathrm{d}N_{t}\,,\quad \widehat{V}_{0}\;=\;I\otimes 
\widehat{1}\,,
\end{equation}%
where the operators $H$ and $C$ act in $\mathcal{H}\otimes \mathcal{F}$ as $%
H\otimes \widehat{1}$ and $C\otimes \widehat{1}$, and $N_{t}=M_{t}+\lambda t%
\widehat{1}$.

The tilde transform $\widetilde{V}_{t}(\tau )$ of $V_{t}(\omega )$ is the
kernel for the process $\widehat{V}_{t}$, and the generating functional $%
\breve{V}_{t}$ coincides with the vacuum conditional expectation $\breve{V}%
_{t}(f)=F_{0}^{\ast }\widehat{V}_{t}^{f}F_{0}$ for $\widehat{V}_{t}^{f}=%
\widehat{V}_{t}(I\otimes W_{t}^{f})$, given by the isometry $F_{0}\eta =\eta
\otimes \varphi _{0}$ of the Hilbert space $\mathcal{H}$ to $\mathcal{H}%
\otimes \mathcal{F}$. It satisfies the ordinary differential equation (\ref%
{gechi}) for each $\eta \in \mathcal{H}$ as $\breve{\chi}_{t}=\breve{V}%
_{t}\eta $ 
\begin{equation}
\displaystyle{\frac{\mathrm{d}}{\mathrm{d}t}}\breve{V}_{t}+\mathrm{i}H\breve{%
V}_{t}\;=\;(C-I)(\lambda ^{-1/2}f(t)+1)\lambda \breve{V}_{t}  \label{dV}
\end{equation}%
with the initial condition $\breve{V}_{0}(f)=I$ for all $f$.

\section{A unitary dilation of the contractive stochastic dynamics}

\setcounter{equation}{0}

A unitary dilation of a causal contractive cocycle $V_{t}(\omega )$ in $%
\mathcal{H}$ (\cite{BelCMP97}, cf.~also \cite{GLSW}) cannot in general be
obtained from a causal unitary stochastic cocycle $U_{t}(\omega
^{0}\,,\,\omega ^{1})$ in the same Hilbert space $\mathcal{H}$ by fixing $%
\omega ^{0}=\omega $ and averaging over additional degrees of randomness $%
\omega ^{1}\in \Omega ^{1}$. (This is not correct unless like in our paper
only classical randomness is considered.) Even a single contraction $C$
might not be represented as a classical mean $\sum_{k}S_{k}\lambda _{k}$ of
a random unitaries $S_{k}$ with some probabilities $\lambda _{k}\geq 0$, $%
\sum_{k}\lambda _{k}=1$. This makes it impossible to find a Hamiltonian
semiclassical dynamics giving the contractive stochastic dynamics of an
unstable quantum system as a result of a reduced description. However, it
can be obtained from a unitary operator $S$ in an extended Hilbert space $%
\mathcal{H}\otimes \mathcal{K}$ as a block-matrix element 
\begin{equation}
C\;=\;(I\otimes e)^{\ast }S(I\otimes e_{0})\,,\quad \parallel e\parallel
=1=\parallel e_{0}\parallel \,.  \label{4.1}
\end{equation}%
Such a dilation describes the contraction $C$ by the probability amplitudes 
\begin{equation}
\langle \eta \otimes e|S(\eta _{0}\otimes e_{0})\rangle \;=\;\langle \eta
|C\eta _{0}\rangle   \label{4.2}
\end{equation}%
of the unitary transitions $\eta _{0}\otimes e_{0}\rightarrow \eta \otimes e$
in $\mathcal{H}\otimes \mathcal{K}$, given by the fixed unital vectors $%
e_{0},\,e\in \mathcal{K}$, as the probability amplitudes of the contractive
transitions $\eta _{0}\rightarrow \eta $. The unitary dilation $S$ of the
contraction $C$ can always be built in the Hilbert space $\mathcal{H}\oplus 
\mathcal{H}$ with the help of two dimensional space $\mathcal{K}=\mathbb{C}%
^{2}$ by realizing $S$ as a Hermitian block-matrix 
\begin{equation}
S=\left( \!\!%
\begin{array}{cc}
S_{0}^{0} & S_{1}^{0} \\ 
S_{0}^{1} & S_{1}^{1}%
\end{array}%
\!\!\right) \,,\quad S_{0}^{0\ast }=S_{0}^{0}\,,\quad S_{0}^{1\ast
}=S_{1}^{0}\,,\quad S_{1}^{0\ast }=S_{0}^{1}\,,\quad S_{1}^{1\ast
}=S_{1}^{1}\,  \label{4.3}
\end{equation}%
with the transition elements $S_{0}^{1}=C$, $S_{1}^{0}=C^{\ast }$ and 
\begin{equation}
S_{0}^{0}\;=\;-(I-C^{\ast }C)^{1/2}\,,\qquad S_{1}^{1}\;=\;(I-CC^{\ast
})^{1/2}\,.  \label{4.4}
\end{equation}%
The unitarity $S^{-1}=S^{\ast }$ of (\ref{4.2}) simply follows from $%
CS_{0}^{0}+S_{1}^{1}C=0$, $S_{0}^{0}C^{\ast }+C^{\ast }S_{1}^{1}=0$. We can
interpret the unit basic vectors $e_{0},e_{1}\in \mathcal{K}$ as the
eigenstates of a quantum meter detecting the death or life of the unstable
particle, correspondingly. In the case $CC^{\ast }=I$ of coisometric $C$ the
unitary operator $S$ describes a transition of the input particle-meter
states $\eta \otimes e_{0}$, $e_{0}=(\delta _{0}^{k})$ to a superposition of
the alive states $\eta ^{1}=C\eta $, corresponding to the vector $%
e_{1}=(\delta _{1}^{k})$, and the dead states $\eta _{0}=-\eta ^{\perp }$,
where $\eta ^{\perp }=\eta -\eta ^{\mbox{\raise1.3pt\hbox{\tiny $\vert$}}}$
is the orthogonal projection to $\eta ^{\mbox{\raise1.3pt\hbox{\tiny
$\vert$}}}=C^{\ast }C\eta $. But the alive states $\eta \otimes e_{1}$
transit only to the states $C^{\ast }\eta \otimes e_{0}$ corresponding to
the exiting of the particle from the detector. Thus, for realization (\ref%
{4.2}) with the input \textquotedblleft vacuum\textquotedblright\ vector $%
e_{0}$, the output vector $e$ in (\ref{4.1}) is the vector $e_{1}$
corresponding to the detection of the unstable particle. The described
unitary dilation of the contraction $C$ suggests a unitary dilation of the
contractive stochastic cocycle $V_{t}$ in the quantum-mechanical sense. It
should be given by a causal unitary cocycle $U_{t}$ in a Hilbert tensor
product $\mathcal{H}\otimes \mathcal{F}_{\bullet }$ with respect to a free
evolution unitary group $T_{t}$ in the additional space $\mathcal{F}%
_{\bullet }$ of an external quantum field, such that 
\begin{equation}
F_{t}^{\ast }(\omega )U_{t}F_{0}\chi =V_{t}(\omega )\chi (\omega )\,,\quad
\forall t\in R_{+}\,,\quad \omega \in \Omega \,.  \label{4.5}
\end{equation}%
Here $\big(F_{t}\big)_{t\geq 0}$ are isometries $\mathcal{H}\otimes
L^{2}(\Omega ,\,P^{\lambda })\longrightarrow \mathcal{H}\otimes \mathcal{F}%
_{\bullet }$ given as $F_{t}(\eta \otimes \varepsilon ^{g})=\eta \otimes
\varphi _{t}^{g}$ by a correspondence $\varepsilon ^{g}\mapsto \varphi
_{t}^{g}$ of the exponential test functions (\ref{epsilong}) of the Hilbert
subspaces $L^{2}(\Omega ,\,P^{\lambda })$ and their representations $\varphi
_{t}^{g}\in \mathcal{F}_{\bullet }$ such that 
\begin{equation}
\parallel \!\varphi _{t}^{g}\!\parallel ^{2}\,\;=\;\int |\varepsilon
^{g}(\omega )|^{2}P^{\lambda }(\mathrm{d}\omega )\;=\;\parallel \!\varphi
_{0}^{g}\!\parallel ^{2}\,,\quad \forall t\in \mathbb{R}_{+}\,.  \label{4.6}
\end{equation}%
As follows from \cite{BelCMP97}, a good candidate for $\mathcal{F}_{\bullet }
$ is the Bosonic Fock space over the tensor product $\mathcal{K}\otimes
L^{2}(\mathbb{R}_{+})$ of two dimensional $\mathcal{K}=\mathbb{C}^{2}$ and
the space of square-integrable functions on $\mathbb{R}_{+}$ such that $%
\mathcal{F}_{\bullet }$ is the Hilbert product $\mathcal{F}_{0}\otimes 
\mathcal{F}_{1}$ of two copies of the Fock space $\mathcal{F}$ isometric to
the probabilistic space $L^{2}(\Omega ,\,P^{\lambda })$ for the Poisson
process on $\mathbb{R}_{+}$. Realizing $\mathcal{F}_{\bullet }$ as the space
of square-integrable tensor-functions $\varphi (\tau )\in \mathcal{K}%
^{\otimes |\tau |}$ of the finite sequences $\tau =\{r_{1},\,\ldots
\,,\,r_{n}\}\subset \mathbb{R}_{+}$, $r_{1}<r_{2}<\,\ldots \,<r_{n}$, we
shall define the isometries $F_{t}$ by the tilde transform (\ref{tildetrans}%
) of $\varepsilon ^{g}\in L^{2}(\Omega ,\,P^{\lambda })$ as 
\begin{equation}
F_{t}(\eta \otimes \varepsilon ^{g})(\tau )\;=\;\eta \otimes \widetilde{%
\varepsilon }^{g}(\tau _{t})e^{\otimes |\tau _{t}|}\otimes \widetilde{%
\varepsilon }^{g}(\tau _{\lbrack t})e_{0}^{\otimes |\tau _{\lbrack t}|}\,,
\label{4.7}
\end{equation}%
where $e,\,e_{0}\in \mathcal{K}$ are unital 2-vectors, and $|\tau _{t}|=\tau
\cap \lbrack 0,\,t)$, $\tau _{\lbrack t}=\tau \cap \lbrack t,\,\infty )$.
The adjoint transform $F_{t}^{\ast }:\mathcal{H}\otimes \mathcal{F}_{\bullet
}\rightarrow \mathcal{H}\otimes L^{2}(\Omega ,\,P^{\lambda })$ can be
written as $\chi _{t}(\omega )=F_{t}^{\ast }(\omega )(\eta \otimes \varphi )$
in terms of $\eta \otimes \big(e^{\otimes |\tau |}|\varphi (\tau )\big)$,
where $e^{\otimes |\tau |}=e^{\otimes |\tau _{t}|}\otimes e_{0}^{\otimes
|\tau _{\lbrack t}|}$, as the stochastic multiple integral; 
\begin{equation}
\chi _{t}(\omega )\;=\;\eta \otimes \int \lambda ^{-|\tau |/2}\big(%
e^{\otimes |\tau |}|\varphi (\tau )\big)\,\mathrm{d}m_{\tau }\,.  \label{4.8}
\end{equation}%
Here we used the canonical decomposition $\mathcal{F}_{\bullet }=\mathcal{F}%
_{\bullet t}\otimes \mathcal{F}_{\bullet \lbrack t}$ to the Fock spaces over
the orthogonal subspaces of square-integrable vector functions $f^{\bullet
}=(f^{k})$ on $[0,t)$ and $[t,\infty )$, respectively. The tensor
multipliers $\mathcal{F}_{\bullet t}$ are increasingly embedded into $%
\mathcal{F}_{\bullet }$ as $\mathcal{F}_{\bullet t}\subset \mathcal{F}%
_{\bullet s}\,\;\forall s>t$ by $\mathcal{F}_{\bullet t}\ni \varphi
_{t}\mapsto \varphi _{t}\otimes \varphi _{0}^{t}$, where $\varphi
_{0}^{t}(\tau )=\delta _{0}^{|\tau _{\lbrack t}|}$ is the vacuum normalized
function of the space $\mathcal{F}_{[t}$.

The Hilbert space $\mathcal{H}_{t}=L_{\mathcal{H}}^{2}(\Omega
,P_{t}^{\lambda })$ of $\mathcal{H}$-valued stochastic causal functions $%
\chi _{t}(\omega )$ with the finite covariance 
\begin{equation}
\parallel \chi _{t}\parallel ^{2}\,\;=\;\int \parallel \chi _{t}(\omega
)\parallel ^{2}P_{t}^{\lambda }(\mathrm{d}\omega )\;=\;\parallel \widetilde{%
\chi }_{t}\parallel ^{2}<\infty   \label{4.9}
\end{equation}%
is causally represented by the initial isometry $F_{0}$ in the space $%
\mathcal{H}\otimes \mathcal{F}_{\bullet t}$ and thus in $\mathcal{H}\otimes 
\mathcal{F}_{\bullet }$ by the tilde transform $F_{0}\chi _{t}=(I\otimes
e_{0}^{\otimes })\widetilde{\chi }_{t}$, where $e_{0}^{\otimes }$ is the
embedding of $\mathcal{F}_{t}$ into $\mathcal{F}_{\bullet t}$, 
\begin{equation}
(I\otimes e_{0}^{\otimes })(\eta \otimes \varphi _{t})=\eta \otimes \varphi
_{t}e_{0}^{\otimes }\,,\quad \varphi _{t}e_{0}^{\otimes }(\tau )=\varphi
_{t}(\tau )e_{0}^{\otimes |\tau |}\,,  \label{4.10}
\end{equation}%
given by the tensor powers of the unit vector $e_{0}\in \mathbb{C}^{2}$.
This representation is an isometry, 
\begin{equation}
\parallel \!F_{0}\chi _{t}\!\parallel ^{2}\;=\;\parallel \!(I\otimes
e_{0}^{\otimes })\widetilde{\chi }_{t}\!\parallel ^{2}\;=\;\parallel \!\chi
_{t}\!\parallel ^{2}  \label{4.11}
\end{equation}%
due to the unitarity of the tilde transform $\chi \mapsto \widetilde{\chi }$%
. The adjoint co-isometry $F_{0}^{\ast }:\mathcal{H}\otimes \mathcal{F}%
_{\bullet }\rightarrow \mathcal{H}\otimes \mathcal{F}$ maps the localized
kernels $\psi _{t}\in \mathcal{H}\otimes \mathcal{F}_{\bullet t}$ to the
stochastic causal functions $\chi _{t}(\omega )=F_{0}^{\ast }(\omega )\psi
_{t}$ given as the stochastic multiple integrals (\ref{iter}) with the
kernels $\widetilde{\chi }_{t}=(I\otimes e_{0}^{\otimes })^{\ast }\psi _{t}$%
: 
\begin{equation}
\widetilde{\chi }_{t}(r_{1}\,,\,\ldots \,,\,r_{n})={(I\otimes e_{0}^{\otimes
n})}^{\ast }\psi _{t}(r_{1}\,,\,\ldots \,,\,r_{n})\,.  \label{4.12}
\end{equation}

Now we can describe the Markov quantum stochastic model of the unitary
dilation (\ref{4.2}), which has been found in \cite{BelCMP97} for the
general CP flows $\phi_{t}^{g}$ over the algebra $\mathcal{B}(\mathcal{H})$
of all bounded operators in $\mathcal{H}$.

Let us assume that the contraction $C$ is dilated as in (\ref{4.1}) to a
unitary operator $S$ in $\mathcal{H}\otimes \mathcal{K}$, say, of the form (%
\ref{4.2}) and take $e_{0}=\big(\delta _{0}^{k}\big)$, $e=\big(\delta
_{1}^{k}\big)\equiv e_{1}$. We shall define the unitary evolution $U_{t}$ in 
$\mathcal{H}\otimes \mathcal{F}_{\bullet }$ by the quantum stochastic
differential equation (QSDE) in the sense of \cite{HudsonParthasarathy} as 
\begin{equation}
\mathrm{d}U_{t}+\mathrm{i}HU_{t}\,\mathrm{d}t=(S_{k}^{i}-I\delta
_{k}^{i})U_{t}\,\mathrm{d}N_{i}^{k}(t)\,,\quad U_{0}=I\otimes \widehat{1}\,,
\label{4.13}
\end{equation}%
where $N_{k}^{i}(t)$ is the quantum stochastic Poisson matrix process of
intensity $\lambda $ given by the canonical integrators in $\mathcal{F}$ as 
\begin{equation}
N_{k}^{i}(t)=\Lambda _{k}^{i}(t)+\sqrt{\lambda }\big(\Lambda
_{-}^{i}(t)\delta _{k}^{0}+\delta _{0}^{i}\Lambda _{k}^{+}(t)\big)+\lambda
\delta _{0}^{i}\delta _{k}^{0}t\widehat{1}\,.  \label{4.14}
\end{equation}%
In the eigenrepresentation of the number process $N=N_{0}^{0}+N_{1}^{1}$ of
total quantum number, the unitary solution to (\ref{4.1}) can be written
similarly to (\ref{product}) as 
\begin{equation}
U_{t}(\omega )_{k_{1}\,\ldots \,k_{n}}^{i_{1}\,\ldots
\,i_{n}}\;=\;S_{k_{n}}^{i_{n}}(t-t_{n})\cdot \,\ldots \,\cdot
S_{k_{1}}^{i_{1}}(t_{2}-t_{1})\mathrm{e}^{-\mathrm{i}Ht_{1}}\,,  \label{4.15}
\end{equation}%
where $S_{k}^{i}(t)=\mathrm{e}^{-\mathrm{i}Ht}S_{k}^{i}$, $n=n_{t}(\omega )$%
, and $\omega \;=\;\{t_{1},\,t_{2},\,\ldots \}$ are the counting points for
the total number process $N$ up to $t$ with the finite numbers $n_{t}(\omega
)=|\omega \cap \lbrack 0,\,t)|$. We shall also define the isometries $F_{t}$
as $(I\otimes J_{t})F_{0}$, where $J_{t}$ is a partial isometry given by the
solution to QSDE 
\begin{equation}
\mathrm{d}J_{t}=(ee_{0}^{\ast }-\delta )_{k}^{i}J_{t}\,\mathrm{d}%
N_{i}^{k}(t)\,,\quad J_{0}=\widehat{1}\,,  \label{4.16}
\end{equation}%
where $e=e_{1}$ if the unitary matrix $S$ is taken in the form (4.3), (4.4).
The equation (\ref{4.16}) has the explicit solution 
\begin{equation}
J_{t}(\omega )\;=\;\big(ee_{0}^{\ast }\big)^{\otimes |\omega _{t}|}\otimes
I^{\otimes |\omega _{\lbrack t}|}\,,  \label{4.17}
\end{equation}%
where $\omega _{t}\cup \omega _{\lbrack t}$ has a finite $\omega _{t}=\omega
\cap \lbrack 0,\,t)$. Note that the family of orthoprojectors $%
I_{t}=J_{t}J_{t}^{\ast }$, that is 
\begin{equation}
I_{t}(\omega )\;=\;\big(ee^{\ast }\big)^{\otimes |\omega _{t}|}\otimes
I^{\otimes |\omega _{\lbrack t}|}  \label{4.18}
\end{equation}%
satisfying 
\begin{equation}
\mathrm{d}I_{t}=(ee^{\ast }-\delta )_{k}^{i}I_{t}\,\mathrm{d}%
N_{i}^{k}(t)\,,\quad I_{0}=I\otimes \widehat{1}\,,  \label{4.19}
\end{equation}%
is decreasing, $I_{t}\leq I_{r}$, $\forall \,t\geq r\in \mathbb{R}_{+}$,
describing the survival events for the detection of an unstable quantum
particle by the time $t\in \mathbb{R}_{+}$. Obviously, $U_{t}$ dilates the
stochastic evolution as $V_{t}(\omega )=U_{t}(\omega )_{0\,\ldots
\,0}^{1\,\ldots \,1}$ coincides with (2.4) if $C=S_{0}^{1}$. Hence, the
unitary evolution $U_{t}\psi _{0}$ of the initial state $\psi _{0}=F_{0}\chi 
$ with any $\chi \in \mathcal{H}\otimes L^{2}(\Omega ,\,P^{\lambda })$
defines the amplitude $\psi _{t}=(I\otimes J_{t}^{\ast })U_{t}\psi _{0}$
normalized to the averaged survival probability (2.6) 
\begin{equation}
\parallel \!\psi _{t}\!\parallel ^{2}\;=\;\langle U_{t}\psi _{0}|(I\otimes
I_{t})U_{t}\psi _{0}\rangle \;=\;\parallel \!F_{t}^{\ast }U_{t}F_{0}\chi
\!\parallel ^{2}\;=\;\parallel \!V_{t}\chi \!\parallel ^{2}\;=\;q^{\lambda
}(t)\,.  \label{4.20}
\end{equation}%
Here we used the adaptedness of the solution $U_{t}$ in the sense 
\begin{equation}
U_{t}(\eta \otimes e_{0}^{\otimes }\varepsilon ^{g})(\tau )\;=\;U_{t}(\eta
\otimes e_{0}^{\otimes }\varepsilon ^{g}(\tau _{t})\otimes e_{0}^{\otimes
|\tau _{\lbrack t}|}\varepsilon ^{g}(\tau _{\lbrack t})\,,  \label{4.21}
\end{equation}%
due to which $(I\otimes J_{t}^{\ast })U_{t}F_{0}=F_{t}^{\ast }U_{t}F_{0}$.

Let us also prove the dilation formula (\ref{4.5}) using the generating
functional method described in the Sec.~3, and find the limit of the unitary
evolution as $\lambda \rightarrow \infty $ and $C=I-\lambda ^{-1}R$. To do
so it is more convenient to use another dilation, given by $e=e_{0}$ and the
equation (\ref{4.13}) with non-Hermitian unitary block matrix 
\begin{equation}
S\;=\;\left( 
\begin{array}{cc}
\!\!C & (I-CC^{\ast })^{1/2}\!\! \\ 
\!\!-(I-C^{\ast }C)^{1/2} & C^{\ast }\!\!%
\end{array}%
\right) \,,\qquad e\;=\;\left( 
\begin{array}{c}
\!\!1\!\! \\ 
\!\!0\!\!%
\end{array}%
\right) \;=\;e_{0}\,.  \label{4.22}
\end{equation}%
We should find an equation for the coherent matrix elements 
\begin{equation}
U_{t}\big(\bar{g}^{\bullet },\,f^{\bullet }\big)\;=\;\big(g^{\otimes
}|U_{t}f^{\otimes }\big)\,/\,\,\exp \big\{\int_{0}^{\infty }\big(g^{\bullet
}(r)|f^{\bullet }(r)\big)\,\mathrm{d}r\big\}  \label{4.23}
\end{equation}%
and compare it with the equation (\ref{dV}) for $\breve{V}_{t}(\bar{g}\dot{+}%
f)=V_{t}(\bar{g},\,f)$, where 
\begin{eqnarray*}
V_{t}(\bar{g},\,f) &=&\int \overline{\varepsilon ^{g}}(\omega )V_{t}(\omega
)\varepsilon ^{f}(\omega )\,P^{\lambda }(\mathrm{d}\omega )\,/\,\,\exp \big\{%
\int_{0}^{\infty }\bar{g}(r)f(r)\,\mathrm{d}r\big\} \\
&=&\int \overline{\varepsilon _{t}^{g}}(\omega )V_{t}(\omega )\bar{%
\varepsilon}_{t}^{f}(\omega )\,P_{t}^{\lambda }(\mathrm{d}\omega
)\,/\,\,\exp \big\{\int_{0}^{\infty }\bar{g}(r)f(r)\,\mathrm{d}r\big\} \\
&=&\breve{V}_{t}(\bar{g}\dot{+}f)\,.
\end{eqnarray*}%
Here we used the multiplication formula (\ref{3.7}) and the independence of $%
\varepsilon _{t}^{\bar{g}\dot{+}f}V_{t}$ and $\varepsilon _{\lbrack t}^{\bar{%
g}\dot{+}f}$, where $\varepsilon _{\lbrack t}^{f}=\varepsilon ^{{f}_{[t}}$ ( 
$f_{[t}(r)=f(r)$ if $r\geq t$, $f_{[t}(r)=0$ if $r<t$).

The equation for $U_{t}(\bar{g}^{\bullet },\,f^{\bullet })$ can be obtained
by the substitution of the independent increments \textrm{d}$N_{i}^{k}(t)$
for the number process (4.14) in (4.13) by their coherent matrix elements 
\begin{equation*}
\big(f^{k}(t)g_{i}(t)+\lambda ^{1/2}\big(\delta
_{0}^{k}g_{i}(t)+f^{k}(t)\delta _{i}^{0}\big)+\lambda \delta _{0}^{k}\delta
_{i}^{0}\big)\mathrm{d}t\widehat{1}\,,
\end{equation*}%
where $g_{i}=\bar{g}^{i}$. Thus we obtain the ordinary differential equation 
\begin{eqnarray}
\lefteqn{\frac{\mathrm{d}}{\mathrm{d}t}U_{t}\big(\bar{g}^{\bullet
},\,f^{\bullet }\big)+\mathrm{i}HU_{t}\big(\bar{g}^{\bullet },\,f^{\bullet }%
\big)=}  \label{4.24} \\
&&\!\!\!\!\!\!\!\!\!\!\!\big[g_{i}(t)(S\!-\!I\delta
)_{k}^{i}f^{k}(t)\!+\!\lambda ^{1/2}\kern-0.5pt\big(g_{i}(t)(S\!-\!I\delta
)_{0}^{i}\!+\!(S\!-\!I\delta )_{k}^{0}f^{k}(t)\big)\!+\!(S_{0}^{0}\!-\!I)%
\lambda \big]U_{t}\kern-0.5pt\big(\bar{g}^{\bullet }\!,f^{\bullet }\kern%
-0.5pt\big).  \notag
\end{eqnarray}%
If $g_{i}=\delta _{i}^{0}\bar{g}$, $f^{k}=\delta _{0}^{k}f$ corresponding to
the embeddings $g^{\bullet }=ge_{0}$, $f^{\bullet }=fe_{0}$, this equation
indeed coincides with the equation (\ref{dV}); 
\begin{eqnarray*}
\lefteqn{\frac{\mathrm{d}}{\mathrm{d}t}V_{t}(\bar{g},\,f)+\mathrm{i}HV_{t}(%
\bar{g},\,f)} \\
&=&\big[\bar{g}(t)(C-I)f(t)+\lambda ^{1/2}\big(\bar{g}(t)(C-I)+(C-I)f(t)\big)%
+(C-I)\lambda \big]V_{t}(\bar{g},\,f) \\
&=&(C-I)\big(\lambda ^{-1/2}(g\dot{+}f)(t)+1\big)\lambda V_{t}(\bar{g}%
,\,f)\,.
\end{eqnarray*}%
Here $V_{t}(\bar{g},f)=U_{t}(\bar{g}\delta _{0}^{\bullet },\,f\delta
_{0}^{\bullet })$ as 
\begin{equation*}
\langle \xi |V_{t}(\bar{g},f)\eta \rangle \!=\!\langle F_{0}(\xi \otimes
\varepsilon ^{g})|U_{t}F_{0}(\eta \otimes \varepsilon ^{f})\rangle \exp
\{-\!\!\int_{0}^{\infty }\!\!\!\bar{g}(r)f(r)\mathrm{d}r\}\!=\!\langle \xi
|U_{t}(\bar{g}e_{0},fe_{0})\eta \rangle 
\end{equation*}%
in the case $F_{t}=F_{0}$ corresponding to $e=e_{0}$.

Now, substituting C by $I-\lambda ^{-1}R$ in (\ref{4.22}) and taking into
account that 
\begin{equation}
S=\left( 
\begin{array}{cc}
\!\!\!I & 0\!\!\! \\ 
\!\!\!0 & I\!\!\!%
\end{array}%
\right) \!+\!\lambda ^{-1/2}\!\left( 
\begin{array}{cc}
\!\!\!0 & (R+R^{\ast })^{1/2}\!\!\! \\ 
\!\!\!-(R+R^{\ast })^{1/2} & 0\!\!\!%
\end{array}%
\right) \!-\!\lambda ^{-1}\!\left( 
\begin{array}{cc}
\!\!\!R & 0\!\!\! \\ 
\!\!\!0 & R^{\ast }\!\!\!\!%
\end{array}%
\right) \!+\!O(\lambda ^{-3/2}),  \label{4.25}
\end{equation}%
let us find the limiting equation (\ref{4.24}) as $\lambda \rightarrow
\infty $. The right hand side of (\ref{4.24}) up to the term of the order $%
\lambda ^{-1/2}$ is written then as 
\begin{equation*}
\big(O(\lambda ^{-1/2})+(R+R^{\ast })^{1/2}f^{1}(t)-g_{1}(t)(R+R^{\ast
})^{1/2}-R\big)U_{t}\big(\bar{g}^{\bullet },\,f^{\bullet }\big)\,,
\end{equation*}%
giving to the equation 
\begin{equation}
\displaystyle{\frac{\mathrm{d}}{\mathrm{d}t}}U_{t}^{0}\big(\bar{g}^{\bullet
},\,f^{\bullet }\big)+\big(R+\mathrm{i}H\big)U_{t}^{0}\big(\bar{g}^{\bullet
},\,f^{\bullet }\big)=\big(R+R^{\ast }\big)^{1/2}U_{t}^{0}\big(\bar{g}%
^{\bullet },\,f^{\bullet }\big)\big(f^{1}(t)-\bar{g}^{1}(t)\big)
\label{4.26}
\end{equation}%
for the limiting $U_{t}^{0}\big(\bar{g}^{\bullet },\,f^{\bullet }\big)%
=\lim_{\lambda \rightarrow \infty }U_{t}\big(\bar{g}^{\bullet },\,f^{\bullet
}\big)$. Equation (\ref{4.26}) corresponds to the diffusion QSDE 
\begin{equation}
\mathrm{d}U_{t}^{0}+\big(R+\mathrm{i}H\big)U_{t}^{0}\mathrm{d}t\,=\,\big(%
R+R^{\ast }\big)^{1/2}U_{t}^{0}\big(\mathrm{d}\Lambda _{-}^{1}-\mathrm{d}%
\Lambda _{1}^{+}\big)\,,\quad U_{0}^{0}=I\otimes \widehat{1}\,,  \label{4.27}
\end{equation}%
for the limiting unitary Markovian evolution, dilating the limiting
nonstochastic contractive evolution (\ref{diff}) for $V_{t}^{0}=\lim_{%
\lambda \rightarrow \infty }V_{t}$. It is driven by the momentum process 
\begin{equation}
P_{t}\;=\;\mathrm{i}\big(\Lambda _{1}^{+}-\Lambda _{-}^{1}\big)(t)\;=\;%
\widehat{w}_{t}  \label{4.28}
\end{equation}%
which is the quantum stochastic representation of the standard Wiener
process $w_{t}$ in the Fock space $\mathcal{F}_{1}$, the copy of the
original Fock space $\mathcal{F}_{0}$ for the representation $N_{0}^{0}=%
\widehat{n}_{t}$ of the Poisson process $n_{t}(\omega )$.

Thus, the quantum stochastic unitary evolution for the unstable particle
dilating the process of weakly random contractions $C=I-\lambda ^{-1}R$ due
to frequent detection of the particle at random times with the rate $\lambda
\rightarrow \infty $ becomes classically stochastic. The time-evolution of
its pure state is described by the It\^{o}-Schr\"{o}dinger equation 
\begin{equation}
\mathrm{d}\psi _{t}^{0}+\big(R+\mathrm{i}H\big)\psi _{t}^{0}\,\mathrm{d}%
t\;=\;\mathrm{i}\big(R+R^{\ast }\big)^{1/2}\,\psi _{t}^{0}\,\mathrm{d}w_{t}
\label{4.29}
\end{equation}%
for $\psi _{t}^{0}(\omega _{1})=U_{t}^{0}(\omega _{1})\eta $. Here $\omega
_{1}$ is an elementary event of the standard Wiener probability space $\big(%
\Omega _{1},\,P_{1}\big)$, $H=H^{\ast }$ is a selfadjoint operator, and $%
R+R^{\ast }\geq 0$ is the rate operator for the contraction semigroup $%
\mathrm{e}^{-Kt}$, $K=R+$\textrm{i}$H$.

Eq.~(\ref{4.29}) provides a new dynamical formulation of the quantum Zeno
effect \cite{Friedman, MisraSudarshan}. The limiting dynamics becomes
reversible (invertible) as the reductions of the increasing rate and
decreasing amplitude can be compensated by field fluctuations given by the
momentum process. Let us stress that the large number limit (\ref{4.29}) of
the unitary dilation of the contractive stochastic dynamics remains
stochastic. To our knowledge, the stochastic dynamics has not been obtained
so far, in a similar context.

\textbf{Acknowledgement} This work supported in part by The State Committee
for Scientific Research, under the project no.~2P03B 129 11. \emph{Dedicated
to R.\thinspace S. Ingarden on the occasion of his 80$^{th}$ birthday}

\end{document}